# The Open Universe Initiative

P. Giommi[1], G. Arrigo[1], U. Barres De Almeida[2], M. De Angelis[1], J. Del Rio Vera[3], S. Di Ciaccio[1], S. Di Pippo[3], S. Iacovoni[1], A.M.T. Pollock[4]

## 1.      Introduction

The almost universal availability of electronic connectivity, web software, and portable devices is bringing about a major revolution: information of all kinds is rapidly becoming accessible to everyone, transforming social, economic and cultural life practically everywhere in the world. Internet technologies represent an unprecedented and extraordinary two-way channel of communication between producers and users of data. For this reason the web is widely recognised as an asset capable of achieving the fundamental goal of transparency of information and of data products, in line with the growing demand for transparency of all goods that are produced with public money.

In the field of space science almost all existing data sets have been produced through public funding, therefore they should be considered a public good and become openly available to anyone at a certain point in time. In particular, high-level calibrated data products, like images, spectra and similar products, should be available in a transparent form, that is usable by all. To ensure a fair scientific reward and to protect the intellectual property of the teams that conceive, design, build and operate the instruments that generate the data, this should happen according to clear rules.

The benefits of openness and transparency have been widely emphasised. They are so large and evident for both users and data providers that even scientific space data generated through private funds should aim at transparency.

Much has been done in recent years, especially in space astronomy, to offer open access, user-friendly platforms and services, demonstrating how natural is the evolution towards a more and more transparent and inclusive ecosystem of tools and services. However, despite the recent progress there is still a considerable degree of unevenness in the services offered by providers of space science data.

Further efforts are necessary to consolidate, standardise and expand services, promoting a significant inspirational data-driven surge in training, education and discovery. Such a process, leading to a much larger level of availability of space science data, should be extended to non-scientific sectors of society.

---

[1] Agenzia Spaziale Italiana, ASI, Rome, Italy
[2] Centro Brasileiro de Pesquisas Físicas, CBPF, Rio de Janeiro, RJ, Brazil
[3] United Nations Office for Outer Space Affairs, UNOOSA, Vienna, Austria
[4] Sheffield University, Sheffield, United Kingdom

To respond to these needs, at the fifty-ninth session of the United Nations Committee on the Peaceful Uses of Outer Space (COPUOS), the Government of Italy, working closely with the Italian Space Agency (ASI), proposed the "Open Universe Initiative".

In this paper we describe the main principles behind the Open Universe, and illustrate some of the features of the first version of a web portal that is under development at ASI.

We also briefly address the potentially very large socio-economical returns of the initiative, whose costs are easily sustainable and certainly marginal with respect to the total investment that is made to produce space science data.

The far reaching vision of the Open Universe Initiative and the potentially global reach, which extends the benefits of space science to large sectors of the society, including emerging and developing Countries, call for a wide international cooperation under the auspices of the United Nations with activities fully integrated into the UN Space2030 agenda.

## 2. The Open Universe Initiative

"Open Universe" is an initiative under the auspices of COPUOS with the objective of stimulating a dramatic increase in the availability and usability of space science data, extending the potential of scientific discovery to new participants in all parts of the world and empowering global educational services.

Following the initial proposal at COPUOS by Italy in 2016 the initiative was included among the activities to be carried out in preparation of the fiftieth anniversary of the first United Nation Conference on the Exploration and Peaceful Uses of Outer Space (UNISPACE+50).

"Open Universe" will ensure that space science data will become gradually more openly available, easily discoverable, free of bureaucratic or administrative barriers, and usable by the widest possible community, from professional space scientists (several thousands of individuals) to citizen scientists (potentially of the order of millions) to the common citizens interested in space science (likely hundreds of millions).

The services delivered by existing space science data producers have significantly improved over time, but are still largely heterogeneous, ranging from basic support reserved to a restricted number of scientists, to open access web sites offering "science-ready" data products, that is high-level calibrated space science data that can be published without further analysis by professionals with suitable knowledge.

"Open Universe" will implement a method of improving the transparency and usability level of the data stored in current space science data archives, and urge the data producers to increase their present efforts so as to extend the usability of space science data to the non-professional community.

After a number of public discussions and international meetings, Open Universe is now being defined in detail under the leadership of the United Nations Office Of Outer Space Affairs (UNOOSA) in close cooperation with the Government of Italy and in collaboration with other participating Countries.

## 2. The ASI Open Universe Portal

In an effort to make progress towards the implementation of the principles put forward by "Open Universe", the Italian Space Agency is developing a prototype web portal for the initiative that aims at becoming a multi-discipline, multi-provider open space data service.

The first version of the ASI Open Universe portal has been released on the occasion of the United Nations/Italy workshop on Open Universe held in Vienna in November 2017 and is openly available at the link openuniverse.asi.it.

The main aims of the portal are:

1. Develop the first prototype of a multi-discipline multi-provider space science web site that aims at data transparency.
2. Concentrate in a single web page the potential of accessing space science data and information from several data archives and related information systems (e.g. catalogues, bibliographic services etc.)
3. Facilitate new types of scientific research based on data-intensive analysis
4. Stimulate discussion among experts and users so as to collect suggestions on how to reach the goals of the "Open Universe" initiative.
5. Help defining the requirements for a new generation of "user-centred" integrated space science data archives that could be used in principle by anyone having access to touch-screen or equivalent technology of the future.
6. Explain and demonstrate the potential of "Open Universe" to the non-space science professionals (e.g. museums, education sector, common citizens).
7. Provide links to a large number of services that give access to space science data services that could be used to evaluate the level of transparency provided.

This web site is not initially meant to be a software system where all space science data can be accessed, viewed and fully analysed in a homogeneous and integrated way. It is rather a web site where a wide range of services developed independently by different space data providers and archive sites can be found in the same place. It is a sort of space science data "shopping mall" where professional scientists and common citizens alike can go, visit and use (shop) the many web services (each clearly identified by its "brand" or developers logos) available next to each other, with the peculiarity that all the services know what the user is looking for as soon as he/she enters the mall.

Fig. 1 shows the front page of the Open Universe portal. It provides general information about the initiative, documentation for the user, several links to space science related sites, and an input area (marked as "User input") where the user can specify his/her requests about astronomical sources, planets, cosmic ray particles etc. The response of the portal following a request is directed to the area marked as "info and output area".

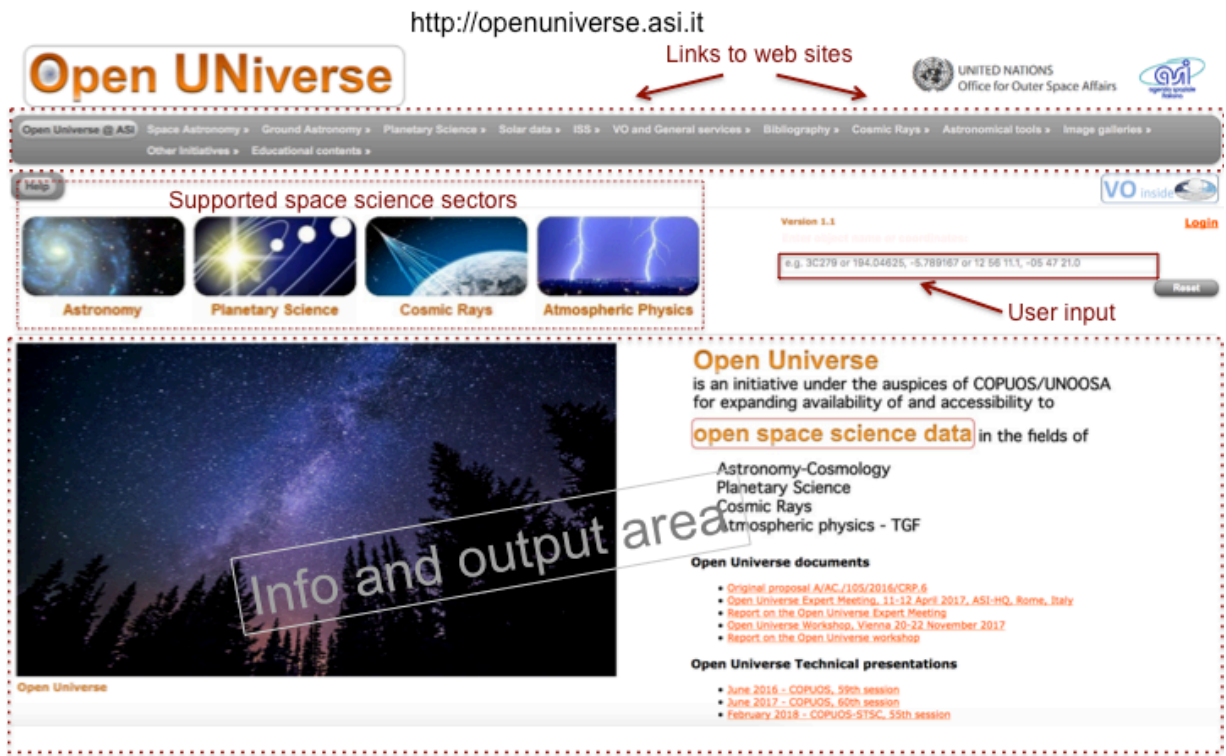

Fig. 1. The front page of the Open Universe portal that can be accessed at the link http://openuniverse.asi.it.

In building this portal we took into account of requirements that maximise usability and accessibility, such as comprehensibility, interactivity, web readiness, and openness. We also cared about avoiding information overload, which is sometimes present in web services in the form of unnecessary or redundant information, and we added synthetic indices.

The grey area in the top part of the portal above the *user input* and highlighted by a red rectangle marked as "links to web sites" gives access to links to the major existing web pages offering data, services or information about space science. Sites are grouped in several different categories, namely: Space astronomy, Ground astronomy, Planetary science, ISS (International Space Station), Virtual Observatory (VO) and General services, Bibliography services  Cosmic Rays, Astronomical web tools, Astronomical image galleries, Open Software, and Educational contents.

Four space science disciplines are supported, namely **Astrophysics, Planetary Science, Cosmic Rays and Atmospheric Physics**. Each discipline corresponds to an icon, as shown in the "supported space science sectors" box of Fig.1. Clicking on one of the four icons activates the corresponding space science area and the "user input" entry point provides appropriate suggestions for the use. For example the Astronomy section is activated then the examples in the input area are set to: 3C279 (for the case of entering the name of an astronomical source) or "194.046, -5.7892" or "12 56 11.1, -05 47 21.0" (for examples of the use of different type of sky coordinates). If the "Planetary science" icon is activated the examples in the entry area are set to "Moon or Mars" etc.

The front page of the portal is where the user has his/her first contact with the system and it is very important that the interaction is as smooth and natural as possible. Great care will be devoted to this requirement. To improve the existing version (V1.1) in the short term we intend to provide a series of video tutorials explaining how to use the system. In the mid/long term we plan to develop new interfaces, based on machine learning techniques that will make the portal as user-centred and effective as possible, in line with our principle that all the existing barriers to data must be as low as possible, ideally removed.

**The Open Universe portal for Astronomy**

As a practical example of the use of the Open Universe portal we show the case of a request for an astronomical source.

By entering in the user's input area the name of a known astronomical source (e.g. M101, or Crab Nebula, Andromeda, 3C273, to name just a few famous astronomical objects) the system activates the astronomy part of the "Open Universe" portal giving access to services offering astronomical data services (Fig. 2). A series of icons appear in the mid part of the page allowing the users to activate services of different types (each identified by a contour of a specific colour[3]), as provided by several existing major web sites. More general information is also provided by services like Wikipedia, NED, and SIMBAD located on the top right part of the portal. In the example of Fig. 2 after entering the name of the galaxy M101 the user clicked on the service "Aladin Lite" from the Centre de Données astronomiques de Strasbourg (CDS), causing the image of M101 to appear in the bottom part of the web page.

---

[3] Blue colour corresponds to services providing astronomical images (from sky surveys at different frequencies), green colour is for astronomical catalogues, red is for data archives, light blue is for sites providing bibliography services, and yellow is used for multi-frequency, data intensive services, that is web sites providing data from several satellites and telescopes.

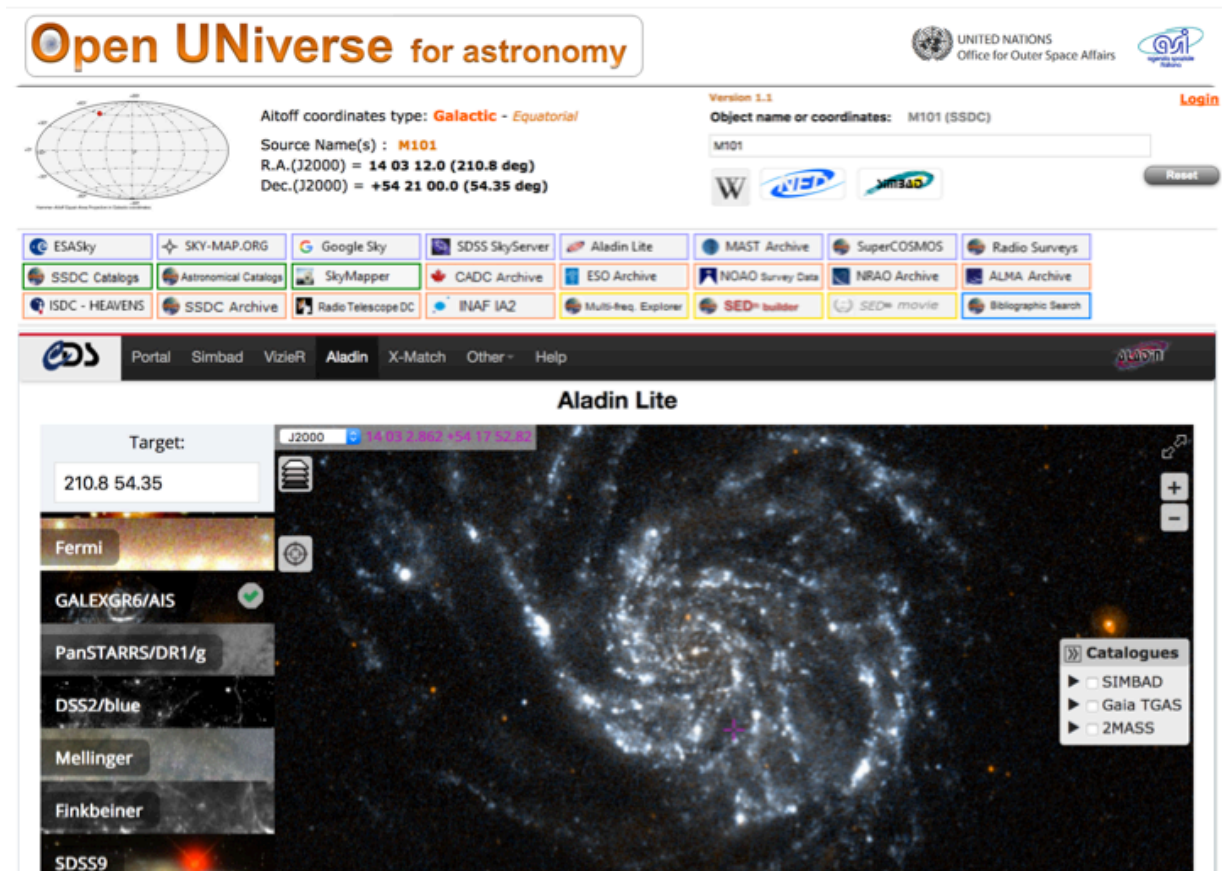

Fig. 2. The Open Universe portal for astronomy: the case of the galaxy M101. The picture shows the image of M101 at ultraviolet frequencies, one of the options provided by the Aladin-Lite tool, a service of CDS in Strasbourg.

## 3. Cost/benefits considerations on the Open Universe Initiative.

**Social and economic benefits**

The Open Universe initiative aims at largely improving the level of transparency, and therefore accessibility and use of space science data both in the scientific community and in non-scientific sectors of the society. In both cases socio-economic benefits may be considerable, in the first case in the form of an increased number of publications in scientific journals, and in the second case in the form of e.g. better education and of a larger discovery potential.

A detailed cost-benefit analysis of the Open Universe initiative is premature and certainly beyond the scope of this paper. In the following we limit ourselves to a few considerations on the social and economic impact of the initiative.

The following is a non-exhaustive list of benefits of the Open Universe initiative.

- Improve services to professional scientists

- Facilitate cross-discipline, multi-experiment and data-intensive research

- Increase the efficiency in the production of knowledge for the same data

- Enable more citizen scientist type activities

- Improve/promote scientific thinking in society

- Inspire young people to scientific careers

- Improve the quality of education

- Extend the potential of scientific discovery to non-professional scientists, in principle to everyone interested in Space Science.

Cost-benefit analysis of research infrastructures is a difficult matter and it is rarely attempted, mostly because of the unpredictability of future economic benefits of science. Nevertheless, it has been recently conducted on some large scientific projects, the most important one being the Large Hadron Collider (LHC), the world's largest particle accelerator, showing that the benefits may be extremely large. For instance, Florio, Forte & Sirtori, (2016[4]), showed that the economic value of the two widely used open access software ROOT and GEANT4 developed and maintained within LHC activities can be estimated to be 2.8 billion Euros.

Expanding the use of space science data to large sectors of society may be achieved by lowering the barriers of usability of space science data to the level of every citizen, (that is offering web-ready or transparent data products) rather than following the more frequently used approach of educating a small fraction of citizens (usually students, researchers or citizen scientists) to be able to perform complex analyses or data processing on low or intermediate level products like raw or un-calibrated data.

The benefits are clearly potentially very large and will depend on how the initiative (and other related activities, e.g. the Virtual Observatory) will evolve.

---

[4] Florio, Forte & Sirtori, Technological forecasting and social Change, 2016, vol. 112, issue C, 38-53. arXiv:1603.00886.

## 4. The role of international cooperation

International cooperation, usually implemented through inter-governmental agreements, plays a large role in the development and the operation of space hardware, especially in very large facilities such as the International Space Station (ISS). Coordination activities, like e.g. the International Exploration Coordination Group (ISECG) are also crucial for the implementation of future highly ambitious space projects, such as human and robotic exploration of the solar system, and the development of bases on the Moon, and Mars. These initiatives clearly hold a large technical, political and economical value.

There are no similarly large and structured examples of international cooperation projects in space science data management, processing and archiving. However, especially in the astronomy sector, some important examples of cooperation emerged in the form of spontaneous aggregation of international research institutions and individuals. Perhaps the most notable one is the cooperation that led to the development of the FITS format. This data format was originally designed in the 1970's for the exchange of radio telescopes images, and over the years, and through the collaboration of experts from space and ground based telescopes operating in all energy bands, evolved into a standard capable of supporting any type of astronomical data type, and endorsed by all major organizations, e.g. NASA, ESA and The International Astronomical Union (IAU).

Today the FITS format is not only a consolidated asset for space science and astronomy, but also a valuable resource that can be used for other purposes. One important example is the use of FITS for the preservation of high-resolution images of ancient documents to ensure that present and future generations will have simple access to the books of the Vatican library.

Another important example of cooperation is the International Virtual Observatory Alliance (IVOA), which, since 2002, focuses on the development of standards and services to the astronomical community to support good data management and interoperability.
The IVOA currently comprises 21 programs from Argentina, Armenia, Australia, Brazil, Canada, Chile, China, Europe, France, Germany, Hungary, India, Italy, Japan, Russia, South Africa, Spain, Ukraine, the United Kingdom, and the United States and an inter-governmental organization (ESA).

Open Universe, with the coordination of UNOOSA, wishes to largely expand the current level of international cooperation in space science data services with large benefits for the scientific community, and other sectors of the society.

## 5. Sustainable costs

What are the costs of achieving the goals of the Open Universe Initiative? Often the answer to this question is "large" or "too large". This is the perception usually expresses by many decision makers and some people operating in the data archiving sector. On close inspection, however, we must note that these costs are certainly negligible compared to the overall investment that has been made to produce space science data, and very small compared to the value that the initiative would generate for our society.

In fact, we argue that the costs associated to Open Universe can be easily sustainable if the following principles are followed.

> For **future space scientific missions**
>
> **Minor adjustments of agencies cost-to-completion models** are necessary to reach the goals of Open Universe: Every year about 15 Billion Euros are spent to produce space science data (not including the sector of Earth observation). By implementing policies at decision making level, that ensure that the final data products meet the Open Universe openness and transparency requirements included the costs of proper data handling, in the overall budget since the very beginning, the cost of reaching transparency would be a tiny percentage of the investment made for the overall project.
>
> For **past, current and future data** the following general principles should be followed to ensure cost-efficiency:
>
> - Avoid duplication of efforts
> - Foster collaboration and coordination among data centres
> - Make use of existing high quality infrastructure and data services
>   (e.g. full use of IVOA standards when possible)
> - Develop innovative tools (e.g. openuniverse.asi.it)
> - Use new paradigms (e.g. distributed analysis?)
> - Take full benefit of new technologies

## 6. Summary and conclusions

Since the advent of the first web-based digital archives offering on-line open astronomical data services in the early nineties, much has been done in the direction of offering space science data to an ever increasing number of users, from the small community of scientists involved in the experiments that produced the data to several thousands of non-specialists researchers. This progress, however, has been strongly discipline dependent, with the astronomy sector leading the way, while other space science disciplines moving at a much lower speed, and in many cases still restricting the use of the expensive data they produce to the small number of scientists belonging to the projects teams.

Much work is still ahead of us to meet these goals. The ASI prototype that has been described above represents a first step along this path. Even in the astronomy sector, today's best digital archives provide in most cases calibrated data and the associated software suitable for higher level scientific analysis that must be carried out by expert users. The rest of the world is still largely excluded from the utilisation of the data. It is too early to predict what is the detailed work that is necessary to achieve all goals of "Open Universe". It is however easy to predict that the associated extra cost is only a tiny fraction of the amount of ≈15 Billion Euros that is spent every year in the world to generate scientific space data.

The Open Universe initiative may play a strategic role in the near future by improving the use of digital technologies to maximize transparency and enhance web interfaces with applications that may go beyond the field of space science and astronomy.

From the time table point of view 2016 has been the year of the proposal to COPUOS[5]; while 2017 was dedicated to ample discussions, with the organization of an expert meeting in Rome[6] and a workshop dedicated to Open Universe open to the international community that took place at the United Nations in Vienna[7].

In 2018 Open Universe will enter a more operational phase, ideally within the UN Space2030 agenda, with the coordination of the United Nations through UNOOSA, in close cooperation with Italy, the scientists and the institutions that collaborated in this initial phase, together with all the UN member States that will support it.

---

[5]http://www.unoosa.org/res/oosadoc/data/documents/2016/aac_1052016crp/aac_1052016crp_6_0_html/AC105_2016_CRP06E.pdf
[6]http://wwwdev.openuniverse.asi.it/documents/ou_documents.php
[7]http://www.unoosa.org/oosa/en/ourwork/psa/schedule/2017/workshop_italy_openuniverse.html